  \documentstyle[aps,prl,twocolumn,epsfig]{revtex}
  
\begin{document}

  \twocolumn[\hsize\textwidth\columnwidth\hsize\csname @twocolumnfalse\endcsname

  \title {Inequivalence of ensembles in a system with long range interactions}

  \author{Julien Barr\'e$^{(a,b)}$, David Mukamel$^{(c)}$, Stefano Ruffo$^{(a,d)}$}

  \address{$^{(a)}$ Dipartimento di Energetica ``Sergio Stecco", Universit\`a
  di Firenze, via s. Marta 3, 50139 Firenze, Italy}

  \address{$^{(b)}$ Ecole Normale Sup\'erieure de Lyon, Laboratoire de Physique,
  46 All\'ee d'Italie, 69364 Lyon Cedex 07, France}

  \address{$^{(c)}$ Department of Physics of Complex Systems, The Weizmann Institute
  of Science, Rehovot 76100, Israel}

  \address{$^{(d)}$ INFM and INFN, Firenze, Italy}

  \date{\today}

  \maketitle

  \begin{abstract}
  We study the global phase diagram of the infinite range
  Blume-Emery-Griffiths model both in the {\it canonical} and in the
  {\it microcanonical} ensembles. The canonical phase diagram is
  known to exhibit first order and continuous transition lines
  separated by a tricritical point. We find that below the
  tricritical point, when the canonical transition is first order,
  the phase diagrams of the two ensembles disagree. In this region
  the microcanonical ensemble exhibits energy ranges with negative
  specific heat and temperature jumps at transition energies.
  These results can be extended to weakly decaying nonintegrable
  interactions.\\

  \end{abstract}

  \vskip 1pc]

  \narrowtext

  Systems in $d$ dimensions with a pairwise interaction potential
  which decays at large distances as $V(r)\sim 1/r^{d+\sigma}$ with
  $-d\leq \sigma \leq 0$, are referred to as {\it nonintegrable}, or
  systems with {\it long range} interactions. Such systems have an ill
  defined thermodynamic limit~\cite{ruelle}. This may be 
  correctly restored by applying the Kac
  prescription~\cite{kac}, within which the potential is rescaled by
  an appropriate, volume dependent, factor which vanishes in the
  thermodynamic limit. However, even within this scheme, the energy
  remains {\it non additive}, i.e. the system cannot be divided into
  independent macroscopic parts, as is usually the case for 
  short range interactions. This fact has no dramatic
  consequences if one is restricted to the {\it canonical} ensemble,
  but it produces striking phenomena in the {\it microcanonical}
  ensemble. For example it may result in a negative specific heat,
  as was first clearly discussed by Lynden-Bell~\cite{lynden} and
  Thirring~\cite{thirring,firstnote}. Indeed, it has been 
  originally observed by Antonov ~\cite{Antonov} that classical 
  gravitational systems
  ($\sigma=-2$, $d=3$) show features of such kind. However, here the
  physical situation is made more complex by the presence of a
  singularity of the interaction potential at short distances. For a
  careful discussion of the statistical mechanics of these systems
  see Ref.~\cite{pad}.

  In the present Letter we consider a simple model for which the
  main features of the phase diagram can be derived analytically
  both within the canonical and the microcanonical ensembles. We
  demonstrate that in the region where the phase transition in the
  canonical ensemble is first order, the two ensembles are not
  equivalent, yielding two distinct phase diagrams. The model we
  consider is the Blume-Emery-Griffiths (BEG) model with infinite
  range interactions ($\sigma=-d$). This is the simplest model known
  to exhibit both continuous and first order transition lines. It is
  defined on a lattice (hence, divergences at short range are
  removed), where each lattice point $i$ is occupied by a spin-$1$
  variable $S_i =1,-1,0$. The Hamiltonian is given by
  \begin{equation}
  H=\Delta\sum_{i=1}^N S_i^2 -\frac{J}{2N}(\sum_{i=1}^N S_i)^2
  \label{BEG}
  \end{equation}
  where $J>0$ is a ferromagnetic coupling constant and $\Delta>0$ 
  controls the energy difference between the magnetic $(S= \pm 1)$ 
  and the non-magnetic $(S=0)$ states. Each spin interacts with 
  every other spin and the coupling constant $J$ is scaled by $1/N$ 
  to ensure {\it extensivity} of the energy. This is just the Kac
  prescription applied to our model. However, this does not entail {\it
  additivity}, in the sense that for a system made of two parts, $X$
  and $Y$, such that $H_{X+Y} = H_X + H_Y + H_{XY}$, the $H_{XY}$
  interaction term never becomes negligible in the thermodynamic
  limit. This property applies to all thermodynamic potentials.

\begin{figure}
  \centering\epsfig{figure=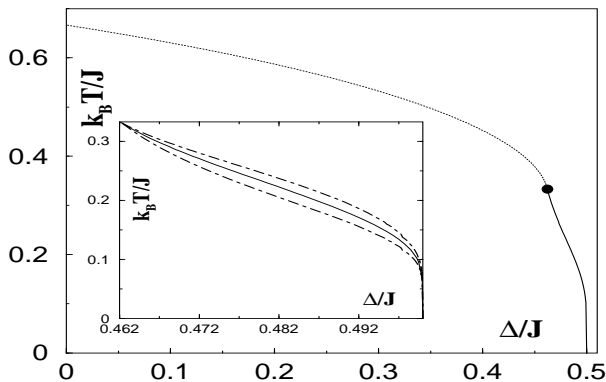,width=5cm,height=8 cm,angle=-90}
  \caption{ Transition lines in the canonical ensemble.
  The critical line (dotted) ends at the tricritical point ($\bullet$),
  where the transition becomes first order (full). The first order
  region is zoomed in the inset, where we show again the canonical
  first order line (full) and the microcanonical transition lines 
  (dot-dashed)}
  \label{fig1}
  \end{figure}

  The canonical phase diagram of this model has been studied in the
  past~\cite{blume}. At $T=0$ the model exhibits a ferromagnetic
  phase for $2\Delta /J <1$ and a non-magnetic phase otherwise. The
  $(T, \Delta)$ phase diagram displays a transition line separating
  the low temperature ferromagnetic phase from the high temperature
  paramagnetic phase (see Fig.\ref{fig1}).
  
  The transition is first order at high $\Delta$ values and becomes
  continuous at low $\Delta$. The critical (second order)
  line is given by
  \begin{equation}
  \label{CriticaLine}
  \beta J = {1 \over 2} e^{\beta \Delta} +1~,
  \end{equation}
  where $\beta = 1/{k_B T}$. The two segments
  of the transition line (high and low $\Delta$) are separated
  by a tricritical point located at $\Delta/J=\ln (4)/3 \simeq 0.4621$,
  $\beta J=3$.
  The first order segment of the transition line is obtained
  numerically by equating the free energies of the ferromagnetic
  and the paramagnetic states.

  We now consider the phase diagram of the BEG model (\ref{BEG}) within
  the microcanonical ensemble. Let $N_+, N_-, N_0$ be the
  number of up, down and zero spin, respectively, in a given
  microscopic configuration. Clearly, $N_+ + N_- + N_0 = N$.
  The energy $E$ of a configuration is obviously a function only
  of $N_+, N_-$ and $N_0$. It does not depend on the specific spatial
  distribution of the spin variables. It is given by
  \begin{equation}
  \label{Energy}
  E=\Delta Q - {J \over {2N}}M^2~,
  \end{equation}
  where $Q=\sum_{i=1}^N S_i^2=N_+ + N_-$ is the quadrupole moment and
  $M=\sum_{i=1}^N S_i=N_+ - N_-$ is the magnetization of the
  configuration. In order to calculate the entropy of a state with
  energy $E$ we note that the number of microscopic configurations
  $\Omega$ compatible with macroscopic occupation numbers
  $N_+, N_-$ and $N_0$ is
  \begin{equation}
  \label{Omega}
  \Omega = {{N!} \over {{N_+!}{N_-!}{N_0!}}}~.
  \end{equation}
  Thus, in the large $N$ limit, the entropy $S=k_B \ln \Omega$
  corresponding to these occupation numbers is given by
  \begin{eqnarray}
  \label{Entropy1}
  S&=&-k_B N [(1-q)\ln (1-q) + {1 \over 2}(q+m) \ln (q+m) \nonumber \\
  &+&{1 \over 2}(q-m) \ln (q-m) - q \ln 2]~,
  \end{eqnarray}
  where $q=Q/N$ and $m = M/N$ are the quadrupole moment and
  magnetization per site, respectively.

  Let $\epsilon = E/{\Delta N}$ be the energy per site, normalized
  by $\Delta$. Equation (\ref{Energy}) may be written as
  \begin{equation}
  \label{q}
  q = \epsilon +Km^2~,
  \end{equation}
  where $K=J/{2\Delta}$. Using this relation, one expresses the
  entropy per site $s=S/(k_B N)$ as a function of $m$ and
  $\epsilon$. Maximizing $s(\epsilon,m)$ with respect to $m$ one
  obtains both the spontaneous  magnetization $m_s(\epsilon)$ and
  the entropy $s(\epsilon)$ of the system for given energy. In order
  to locate the continuous transition line between the paramagnetic
  and the ferromagnetic phases we expand $s(\epsilon,m)$ in powers
  of $m$. This expansion takes the form
  \begin{equation}
  \label{EntropyExpansion}
  s=s_0 +Am^2 + Bm^4 + O(m^6)~,
  \end{equation}
  where $s_0 \equiv s(\epsilon,m=0)$ is the entropy at zero
  magnetization
  \begin{equation}\label{s0}
  s_0=-(1-\epsilon)\ln(1-\epsilon) - \epsilon \ln \epsilon +
  \epsilon \ln 2~,
  \end{equation}
  and $A$ and $B$ are the expansion coefficients
  \begin{eqnarray}\label{AB}
  A&=&-K \ln {\epsilon \over {2(1-\epsilon)}} -{1 \over {2 \epsilon}}
  \nonumber \\
  B&=&-{K^2 \over {2 \epsilon(1-\epsilon)}}+{K \over {2
  \epsilon^2}}-{1\over {12 \epsilon^3}}~.
  \end{eqnarray}
  In the paramagnetic phase both $A$ and $B$ are negative, and the
  entropy is maximized by $m=0$. The continuous transition to the
  ferromagnetic phase takes place at $A=0$ for $B<0$. In order to
  obtain the critical line in the $(T,\Delta)$ plane we note that
  the energy $\epsilon$ is related to the temperature by the usual
  thermodynamic relation
  \begin{equation}
  \frac{\Delta}{k_B T}= \frac{\partial s}{\partial \epsilon}~.
  \label{temp}
  \end{equation}
  Making use of the fact that the magnetization $m$ vanishes on
  the critical line one obtains
  \begin{equation}
  \label{Temperature}
  \frac{\Delta}{k_B T} = \ln {{2(1- \epsilon)} \over \epsilon}~.
  \end{equation}
  This relation, together with the equation $A=0$, yields the
  following expression for the critical line
  \begin{equation}
  \label{MicroCritical}
  2 \bar\beta K ={1 \over 2} e^{\bar\beta} +1~,
  \end{equation}
  where $\bar\beta \equiv \beta \Delta$.
  Equivalently, this expression may be written as
  $\bar\beta K = 1 /{2\epsilon}$. The microcanonical critical line
  thus coincides with the critical line (\ref{CriticaLine}) obtained
  for the canonical ensemble.
  The tricritical point of the microcanonical ensemble is obtained
  at $A=B=0$. Combining these equations with (\ref{Temperature})
  one finds that at the tricritical point $\bar\beta$ satisfies
  \begin{equation}
  \label{MicroTricritical}
  {1 \over {8 {\bar\beta}^2}}{{e^{\bar\beta} +2}\over {e^{\bar\beta}}} -{1\over {4
  {\bar\beta}}} + {1 \over 12} =0~.
  \end{equation}
  Equations (\ref{MicroCritical} ,\ref{MicroTricritical}) yield a
  tricritical point at $K \simeq 1.0813$, $\bar\beta \simeq 1.3998$.
  This has to be compared with the canonical tricritical point
  located at $K=3/ \ln (16) \simeq 1.0820$, $\bar\beta = \ln (4) \simeq
  1.3995$. It is evident that the two points, although very close to
  each other, do not coincide and the microcanonical critical line
  extends beyond the canonical one. In the region between the two
  tricritical points, the canonical ensemble yields a first order
  transition at a higher temperature, while in the microcanonical 
  ensemble the transition is continuous.

\begin{figure}
  \centering\epsfig{figure=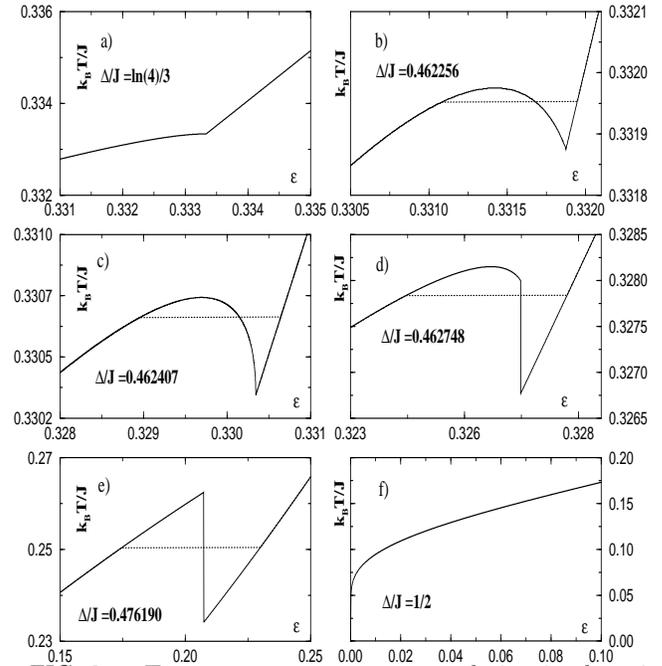,width=9cm,height=8.5cm,angle=-90}
  \caption{ Temperature versus energy relation in the microcanonical
  ensemble for different values of $\Delta$. The dotted horizontal line 
  in some of the plots is the Maxwell construction in the canonical 
  ensemble and identifies the canonical first order transition temperature.}
  \label{fig2}
  \end{figure}

\begin{figure}
  \centering\epsfig{figure=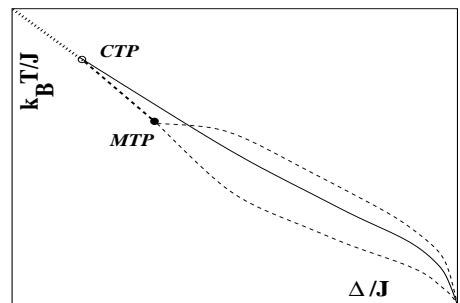,width=6cm,height=4 cm,angle=0}
  \caption{A schematic representation of the phase diagram, where
  we expand the region around the canonical (CTP) and the microcanonical
  (MTP) tricritical points. The second order line, common
  to both ensembles, is dotted, the first
  order canonical transition line is full and the microcanonical transition
  lines are dashed (with the bold dashed line representing a continuous
  transition).}
  \label{fig3}
  \end{figure}

  To study the microcanonical phase diagram we consider the
  temperature-energy relation $T(\epsilon)$. This curve has two
  branches: a high energy branch (\ref{Temperature}) corresponding
  to $m=0$, and a low energy branch obtained from (\ref{temp}) using
  the spontaneous magnetization $m_s(\epsilon)$. At the intersection
  point of the two branches the two entropies become equal. In
  Fig.~\ref{fig2} we display the $T(\epsilon)$ curve for increasing
  values of $\Delta$. For $\Delta /J =\ln(4)/3)$, corresponding to
  the canonical tricritical point, the lower branch of the curve has
  a zero slope at the intersection point (Fig.~\ref{fig2}a). Thus,
  the specific heat of the ordered phase diverges at this point.
  Increasing $\Delta$ to the region between the two tricritical
  points a {\it negative specific heat} in the microcanonical
  ensemble first arises ($\partial T/\partial \epsilon <0$), see
  Fig.~\ref{fig2}b. At the microcanonical tricritical point $\Delta$
  the derivative $\partial T/\partial \epsilon$ of the lower branch
  diverges at the transition point, yielding a vanishing specific
  heat (Fig.~\ref{fig2}c). For larger values of $\Delta$ a jump in
  the temperature appears at the transition energy
  (Fig.~\ref{fig2}d). The lower temperature corresponds to the $m=0$
  solution (\ref{Temperature}) and the upper one is given by $\exp
  (\bar \beta)= 2(1-q^*)/\sqrt{(q^*)^2-(m^*)^2}$, where $m^*,q^*$
  are the values of the order parameters of the ferromagnetic state at the transition
  energy. The negative specific heat branch disappears at larger
  values of $\Delta$, leaving just a temperature jump (see
  Fig.~\ref{fig2}e). In the $\Delta/J \to 1/2$ limit the low temperature
  branch, corresponding to $q=m=1$ in the limit, shrinks to zero
  and the $m=0$ branch (\ref{Temperature}) describes the full
  energy range (Fig.~\ref{fig2}f). In the inset of Fig.~\ref{fig1} we report the
  transition temperatures in the microcanical ensemble against
  $\Delta/J$ for both the $m=0$ (lower dot-dashed line) and the $m\neq0$
  solutions (upper dot-dashed line). The lines are drawn starting at the
  canonical tricritical point. The region between the two
  tricritical points is too small to be appreciated in
  the figure. A schematic phase diagram in the
  first order region is given in Fig.~\ref{fig3}, where we ficticiously
  expand the region of the tricritical points. Note that the
  canonical first order line necessarily crosses the upper
  microcanonical transition line at some point.

  That such unusual effects in the microcanonical ensemble are
  associated with a first order canonical phase transition was also
  suggested in Ref.~\cite{gross1}. These authors discuss also short
  range interactions, for which such features are produced by finite
  size effects.

  As usual for mean-field models, one can express the free energy
  $f(T,m)$ in the canonical ensemble as a function of $T$ and $m$.
  The spontaneous magnetization $m_s(T)$, the temperature-energy
  relation $T(\epsilon)$ and the free energy $f(T)$ may be obtained
  by minimizing $f(T,m)$ with respect to $m$ and using well known
  thermodynamic relations. We now note that the negative specific
  heat branch of the microcanonical ensemble corresponds to a local
  {\it maximum} of the free energy $f(T,m)$ with respect to $m$.

  This result can indeed be derived on quite a general ground. It is
  easy to show that an extremum of $f(T,m)$ corresponds to an
  extremum of $s(\epsilon,m)$ with respect to $m$. Indeed,
  the free energy $f(T,m)$ may be obtained by
  minimizing $\tilde{f}(T,\epsilon,m)=\Delta\epsilon -s/\beta$ with
  respect to $\epsilon$, keeping $T$ and $m$ fixed. This
  minimization yields the usual temperature-entropy relation
  (\ref{temp}). Further minimizing $\tilde{f}(T,\epsilon,m)$ with
  respect to $m$ yields the result that $\partial f(T,m)/\partial m$
  and $\partial s(\epsilon,m)/\partial m$ are proportional to each
  other and thus vanish together. It can be shown~\cite{BMR} by studying the
  second derivatives that when the stationary point of
  $\tilde{f}(T,\epsilon,m)$ with respect to $\epsilon$ and $m$ is a
  saddle point, the resulting entropy exhibits a negative specific
  heat. As a consequence, we can recover the full
  microcanonical solution by studying the stationary points of the
  function $\tilde{f}(T,\epsilon,m)$. However, this function is
  not typically available for non mean-field models.

  The relevant features of the BEG model with infinite range couplings 
  persist also for {\it nonintegrable} interactions. In order to
  investigate this point, we introduce a generalization of the BEG
  model given by the Hamiltonian
  \begin {equation}
  H = \Delta \sum_{i=1}^{N} S_i^2 -\frac{J}{\tilde{N}}\sum_{i>j}
  \frac{S_iS_j}{r_{ij}^{\alpha}}~,
  \label{alpha}
  \end{equation}
  where $r_{ij}$ is the distance on a 1D lattice between spins at
  sites $i$ and $j$. The interactions are non integrable for $\alpha
  \le 1$. The normalization 
  $\tilde{N}=2^\alpha (N^{1-\alpha}-1)/(1-\alpha)$ ensures
  that the energy is extensive. Models of this kind have been previously
  introduced by other authors, and studied within the canonical
  ensemble~\cite{tsallis}. We apply periodic boundary conditions (p.b.c.),
  for which the model is more easily tractable, and we then take $r_{ij}$ to be 
  the smallest of the two distances compatible with p.b.c.. 
  The interaction matrix $({r_{ij})^{-\alpha}}/\tilde{N}$ can
  be exactly diagonalized, which allows to solve model (\ref{alpha})
  in the canonical ensemble. When appropriately rescaled thermodynamic
  quantities are chosen, the solution is the same as for the $\alpha=0$ case (as it
  happens for the models studied in Ref.~\cite{tsallis}).
  Moreover, using a Fourier representation of Hamiltonian (\ref{alpha})
  and considering only the long wavelength components, it is possible
  to obtain an approximate expression for the entropy in the microcanonical
  ensemble~\cite{BMR}. Maximizing this expression at fixed energy, we find
  that the $\alpha = 0$ microcanonical solution is also left unchanged. 
  Since both the canonical and the microcanonical solutions are not modified, 
  we conclude that ensemble inequivalence persists for the slowly 
  decreasing case $\alpha <1$. Details of this analysis will be reported
  elsewhere~\cite{BMR}.

  In summary, we have compared the {\it canonical} with the {\it
  microcanonical} solutions of the infinite range
  Blume-Emery-Griffiths model. We find that the global phase
  diagrams are different in the two ensembles. Although they are found to be
  the same in the domain where the canonical transition is
  continuous, they differ from each other when the canonical
  transition is first order. {\it Negative specific heat} and
  {\it temperature jumps} at the transition energy are found in the
  microcanonical ensemble. These results generalize those of
  Ref.~\cite{thirring} in the context of a simple model, where by
  varying a single parameter one can observe a variety of possible
  features of the phase diagram. Moreover, we are able to understand
  the role played by the constraint of fixing the energy in the
  microcanonical ensemble, which produces a stabilization of
  canonically unstable solutions. In the phase coexistence region,
  the unusual microcanonical thermodynamic properties should result
  in some peculiar dynamical behavior, as has been observed in studies of
  a different mean-field model with continuous
  variables~\cite{antoni}. Our results for the BEG model are not
  limited to the infinite range case, but can be extended to weakly
  decaying {\it nonintegrable} interactions.

  We thank L. Casetti, E.G.D. Cohen, D.H.E. Gross, M.C. Firpo, O. Fliegans,
  F. Leyvraz, M. Pettini, E. Votyakov for stimulating discussions.
  A special thank goes to A. Torcini who collaborated with us in a first
  phase of this project.
  This work is supported by the INFM-PAIS project {\it
  Equilibrium and nonequilibrium dynamics in condensed matter}.

\begin{thebibliography}{99}
\bibitem{firstnote} One should however note that the
  canonical ensemble has no foundation if it cannot be derived
  from the microcanonical, and this is possible only for additive
  systems~\cite{gross}.
\end{thebibliography}
  \end{document}